\documentclass[twocolumn,showpacs,preprintnumbers,amsmath,amssymb]{revtex4}

\usepackage{graphicx}
\usepackage{dcolumn}
\usepackage{bm}

\begin{document}


\title{Room-temperature detection of spin accumulation in silicon across Schottky tunnel barriers using a MOSFET structure}

\author{K. Hamaya,$^{1,2}$\footnote{E-mail: hamaya@ed.kyushu-u.ac.jp} Y. Ando,$^{1}$\footnote{Present address : Graduate School of Engineering Science, Osaka University, Toyonaka, Osaka 560-8531, Japan} K. Masaki,$^{1}$ Y. Maeda,$^{1}$  Y. Fujita,$^{1}$ S. Yamada,$^{1}$ K. Sawano,$^{3}$ and M. Miyao$^{1}$}

\affiliation{$^{1}$Department of Electronics, Kyushu University, 744 Motooka, Fukuoka 819-0395, Japan}%
\affiliation{$^{2}$PRESTO, Japan Science and Technology Agency, Sanbancho, Tokyo 102-0075, Japan}%
\affiliation{$^{3}$Advanced Research Laboratories, Tokyo City University, 8-15-1 Todoroki, Tokyo 158-0082, Japan}

\date{\today}
\begin{abstract}
Using a metal-oxide-semiconductor field effect transistor (MOSFET) structure with a high-quality CoFe/$n^{+}$-Si contact, we systematically study spin injection and spin accumulation in a nondegenerated Si channel with a doping density of $\sim$ 4.5 $\times$10$^{15}$cm$^{-3}$ at room temperature. By applying the gate voltage ($V_\text{G}$) to the channel, we obtain sufficient bias currents ($I_\text{Bias}$) for creating spin accumulation in the channel and observe clear spin-accumulation signals even at room temperature. Whereas the magnitude of the spin signals is enhanced by increasing $I_\text{Bias}$, it is reduced by increasing $V_\text{G}$ interestingly. These features can be understood within the framework of the conventional spin diffusion model. As a result, a room-temperature spin injection technique for the nondegenerated Si channel without using insulating tunnel barriers is established, which indicates a technological progress for Si-based spintronic applications with gate electrodes.
 
\end{abstract}
\maketitle

\section{Introduction}
Owing to the future intrinsic limits of the shrinking of silicon (Si)-based conventional complementary metal-oxide-semiconductor (CMOS) transistors, one should develop novel devices with additional functionalities. In particular, spin-based electronics (spintronics) are expected to markedly improve device performances because of its nonvolatility, reconstructibility, low power consumption, and so forth.\cite{Wolf,Igor,Takanashi,Taniyama} To combine the spintronics with Si-based semiconductor industry, it will become important to explore spintronic technologies compatible with Si.\cite{Jansen1} 

Fortunately, Si has been predicted to be a semiconductor with enhanced spin lifetime and spin transport length due to its low spin-orbit scattering and lattice inversion symmetry.\cite{Igor} Focusing on these characteristics, Appelbaum {\it et al}. demonstrated spin transport across more than micrometer-order Si channels in spin-dependent ballistic hot-electron transport devices.\cite{Appelbaum,Appelbaum1} 
Since they extracted lots of useful data at low temperatures, the phonon-induced spin relaxation in conduction band in Si was also discussed in theories.\cite{Cheng,Dery2} Recently, even at room temperature, electrical detections of spin accumulation with the three-terminal technique \cite{Jansen2,Jonker2} and of spin transport with the four-terminal technique\cite{Suzuki} were demonstrated in lateral devices with degenerated channels. 

For developing spin metal-oxide-semiconductor field-effect transistors (MOSFETs) suggested,\cite{Sugahara,Dery1} there are still many issues. Source and drain (S/D) contacts with insulating tunnel barriers have so far been utilized for electrical spin injection techniques for Si devices\cite{Appelbaum,Appelbaum1,Jansen2,Suzuki,Jonker1,Sasaki2,Jansen3,Ishikawa,Jeon,Jonker2} because the tunnel barriers provide the solution of the impedance mismatch\cite{Rashba,Fert,Takahashi} and of the silicidation reaction\cite{Huang} between ferromagnetic metal (FM) and semiconductors (SC). As a result, highly efficient spin injection into Si \cite{Huang} and long-distance lateral spin transport in a nondegerated Si channel,\cite{Jang1} a Si/SiO$_{2}$ interface,\cite{Jang2,Li2} and an undoped Si channel\cite{Li} were demonstrated in hot-electron transport devices with Al$_{2}$O$_{3}$ tunnel barriers at low temperatures. However, the use of the insulating tunnel barriers basically results in high contact resistance at the interface. For scalable spin MOSFETs with ultralow power consumption, reducing the parasitic resistance between S/D contacts will be strongly required. Thus, we aim to simultaneously realize highly efficient spin injection and low-resistance S/D contacts in Si-based devices. 

To date, we have individually explored two important technologies, high-quality epitaxial growth of FM alloys including half-metallic Heusler alloys on Si\cite{Hamaya,Yamada,Maeda,Oki,Tanikawa} and spin injection into Si across the Schottky-tunnel barrier.\cite{Ando1,Ando2,Ando3,Ando4,Ando5,HamayaJJAP} If we simultaneously realize the above two technologies in a single device, we will be able to open a way for highly efficient spin injection and detection with low parasitic resistance in the future. Recently, we developed electrical detection of spin-accumulation signals in Si with a metal-oxide-semiconductor field-effect transistor (MOSFET) structure with a high-quality CoFe/$n^{+}$-Si contact.\cite{Ando4} Interestingly, the spin signals can be modulated by applying the gate voltage even at room temperature.\cite{Ando4} However, the precise mechanism of the gate-modulated spin signals has not yet been discussed in detail.

In this paper, to discuss the gate-modulated spin signals in detail, we systematically study room-temperature spin injection, spin detection, and spin accumulation in non-degenerated Si channels using a metal-oxide-semiconductor field effect transistor (MOSFET) structure with a high-quality CoFe/$n^{+}$-Si contact. As a result, all the spin signals detected by the three-terminal Hanle-effect measurements for our device can quantitatively be understood within the framework of the standard spin diffusion model.\cite{Jansen1} To precisely discuss the magnitude of spin signals, the correlation between tunnel spin polarization  and resistivity of the channel with the change in the gate voltage should be considered in fabricated devices. This study shows a technological progress for Si-based spintronic applications with non-degenerated Si channels and without using insulating tunnel barriers. 
\begin{figure}[t]
\includegraphics[width=7.5cm]{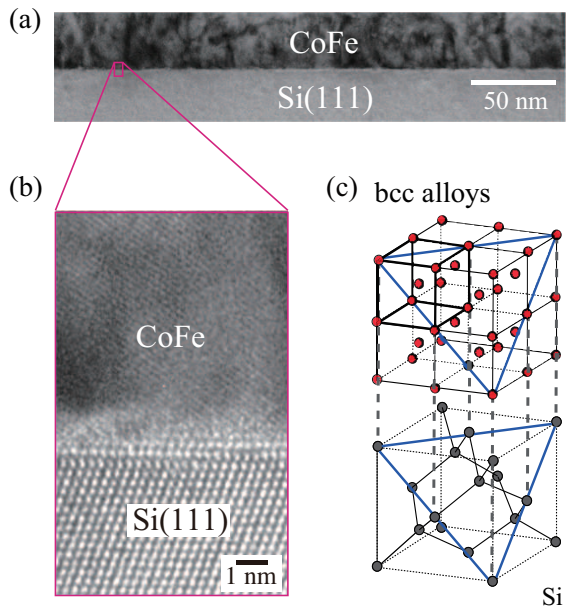}
\caption{(Color online) Cross-sectional (a) low-magnification and (b) high-resolution TEM images of CoFe/Si(111) interface. (c) Crystal structures of bcc-type FM and Si. }
\end{figure} 

\section{High-quality Schottky tunnel contact}
Since the lattice mismatch between some of bcc-type FMs (0.564 $\sim$ 0.570 nm) and Si (0.543 nm) is less than 5\%, we can expect to realize epitaxial growth of the FM films on Si. Until now, we have explored the growth of Heusler alloys and CoFe ones on Si(111).\cite{Hamaya,Yamada,Maeda,Oki} As a result, the bcc-type FM films were successfully grown on Si at 60 $\sim$ 200 $^{\circ}$C using low-temperature molecular beam epitaxy (MBE).\cite{Hamaya,Yamada,Maeda,Oki,Tanikawa} 

Since CoFe alloys are conventional spin injector and detector, we will present a growth process in detail. First, an $n$-type Si wafer with (111) orientation was used as the substrate. After cleaning the substrates with an aqueous HF solution (HF : H$_{2}$O = 1 : 40), we conducted a heat treatment at 450 $^{\circ}$C for 20 min in an MBE chamber with a base pressure of 2 $\times$ 10$^{-9}$ Torr. Prior to the growth of CoFe films, the substrate temperature was reduced to less than 60 $^{\circ}$C. Using Knudsen cells, we co-evaporated Co and Fe, leading to precisely tuning the chemical composition.\cite{Maeda} During the growth, two-dimensional epitaxial growth was confirmed by the observation of reflection high-energy electron diffraction (RHEED) patterns.\cite{Maeda} Figure 1(a) shows a cross-sectional transmission electron microscopy (TEM) image of Co$_{45}$Fe$_{55}$ on Si(111). We can see almost no roughness at the CoFe/Si interface and no interfacial reaction layer at the entire region observed. A high-resolution TEM image near the interface is shown in Fig. 1(b). Almost atomically smooth interface can be seen and single crystalline CoFe alloys can be achieved. We have already examined the influence of atomic composition on the interfacial reaction for Co$_{100-x}$Fe$_{x}$ alloys. As a result, we obtained high-quality films on Si(111) from ${x} =$25 to 55.\cite{Maeda} 

Why did we demonstrate such high-quality epitaxial growth without silicidation reactions ? We now consider the effect of crystal growth with atomically matched heterointerface at the (111) plane. Figure 1(c) illustrates the crystal structures for bcc-type FM alloys and Si. Looking at the (111) plane, denoted by triangles, there is very good atomic arrangement matching between FM and Si. We speculate that this special correlation between FM and Si can realize lowering crystallization energy. Actually, we have achieved high-quality epitaxial growth of Heusler alloys on Si.\cite{Hamaya,Yamada,Oki} This special condition can also be used for the growth of FM alloys on Ge.\cite{Ueda,Ando6,Hamaya1,Hamaya2,Kasahara,Yamada2} Recently, room-temperature structural ordering of a Heusler alloy on Ge\cite{Yamada3} and even the growth of Ge films on a FM alloy can be demonstrated surprisingly by using the (111) plane.\cite{Yamada4}
\begin{figure}[t]
\includegraphics[width=8cm]{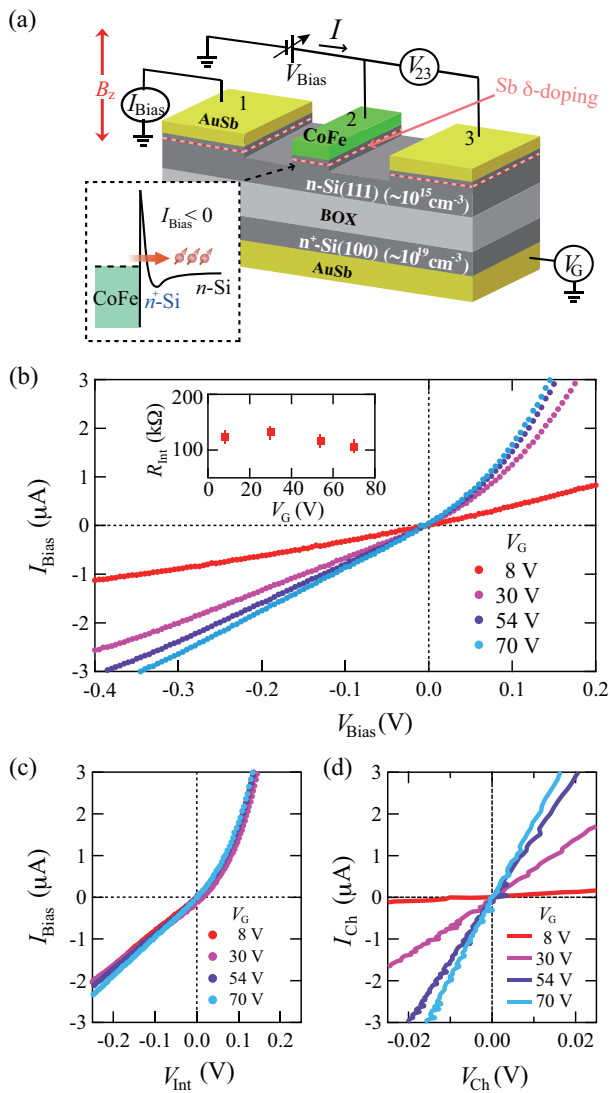}
\caption{(Color online) (a) Schematic diagram of a Si-MOSFET structure with a CoFe/$n^{+}$-Si Schottky-tunnel contact fabricated. (b) $I_\text{Bias} -V_\text{Bias}$ characteristics for various $V_\text{G}$ at room temperature. The inset shows $R_\text{Int}$ vs $V_\text{G}$ at $V_\text{Int} =$ -0.2 V. (c) $I$$_\text{Bias}-$$V_\text{Int}$ and (d) $I$$_\text{Ch}-$$V_\text{Ch}$ curves for various $V_\text{G}$ at room temperature.}
\end{figure}

\section{MOSFET for spin injection and detection}
So far, the back-gated Si spin devices with hot-electron transport have been investigated.\cite{Jang2,Li2} Unlike these,\cite{Jang2,Li2} we focus on a conventional MOSFET device for drift-diffusion spin transport. To fabricate a MOSFET structure with non-degenerated Si channels and Schottky-tunnel contacts, we utilized a CoFe alloy and a (111)-oriented Silicon On Insulator (SOI) substrate. Using low-temperature MBE (LT-MBE) techniques described above, the CoFe epitaxial layers with a thickness of $\sim$ 10 nm were grown on (111)-SOI at $\sim$ 25 $^{\circ}$C,\cite{Maeda} where the thicknesses of the SOI and buried oxide (BOX) layers were about 75 and 200 nm, respectively. Since the doping density (P) of the SOI layer is $\sim$ 4.5 $\times$ 10$^{15}$cm$^{-3}$ (1 $\sim$ 5 $\Omega$cm) at room temperature, the used Si channel is a non-degenerated semiconductor. To obtain tunneling conduction of electrons for spin injection and detection, $n^{+}$-Si layer was inserted between CoFe and SOI by a combination of the Si epitaxy using an MBE process with an Sb $\delta$-doping technique,\cite{Nakagawa,Miyao} where the doping density of Sb atoms is $\sim$1 $\times$ 10$^{19}$cm$^{-3}$.

The Sb $\delta$-doped $n$$^{+}$-Si layer on the channel region was removed by the Ar$^{+}$ ion milling. An ohmic contact (AuSb) for backside heavily doped Si was formed at less than 300 $^{\circ}$C. Conventional processes with electron-beam lithography, Ar$^{+}$ ion milling, and reactive ion etching were used to fabricate three-terminal lateral devices with a backside gate electrode, illustrated in Fig. 2(a). The CoFe/$n^{+}$-Si contact (contact 2) and AuSb ohmic contacts (contact 1 and 3) have lateral dimensions of 1 $\times$ 100 $\mu$m$^{2}$ and 100 $\times$ 100 $\mu$m$^{2}$, respectively. The distance between the contacts 2 and 1 or 3 is $\sim$ 30 $\mu$m.

Using this device structure, we firstly confirm a MOSFET operation at room temperature. With increasing gate voltage ($V_\text{G}$), the bias current ($I$$_\text{Bias}$) value gradually increases with respect to the bias voltage ($V$$_\text{Bias}$) [Fig. 2(b)]. This means that the conduction channel is formed from the vicinity of the interface between SOI and BOX by the $V_\text{G}$ applications, clearly indicating that this device can operate as a MOSFET. Figure 2(c) show $I$$_\text{Bias}-$$V_\text{Int}$ curves for various $V_\text{G}$, where $V_\text{Int}$ is the bias voltage applied to the CoFe/$n^{+}$-Si interface. Here $V_\text{Int}$ is $V_\text{23}$ in the three-terminal measurement. We note that there are almost no changes in the $I$$_\text{Bias}-$$V_\text{Int}$ curves despite applying $V_\text{G}$. Reflecting this feature, we find almost constant interface resistance ($R_\text{Int}$) values (105 $\sim$ 120 k$\Omega$), being independent on $V_\text{G}$, in the inset of Fig. 2(b). The resistance area product of the CoFe/$n^{+}$-Si contact is estimated to be $\sim$ 10$^{7}$ $\Omega$$\mu$m$^{2}$, which is sufficient large value for the spin injection and detection in nondegenerated Si channels with a carrier density of $\sim$ 10$^{15}$cm$^{-3}$.\cite{Jansen1,Fert} 

We also show $I$$_\text{Ch}-$$V_\text{Ch}$ curves for various $V_\text{G}$, where $V_\text{Ch}$ is the bias voltages applied to the Si channel. Here $V$$_\text{Ch}$ is the subtracted value from $V$$_\text{Bias}$ to $V$$_\text{23}$. Contrary to Fig. 2(c), we can clearly see the significant modulation of $I$$_\text{Ch}$ values by applying $V_\text{G}$. That is, we can recognize that the $I$$_\text{Bias}-$$V$$_\text{Bias}$ characteristic shown in Fig. 2(b) is largely dominated by the change in the characteristics of the Si channel. By applying $V_\text{G}$, we can obtain sufficient large current flows of several $\mu$A for spin injection from the CoFe contact to the Si channel. 

\section{Spin accumulation in nondegenerated Si}
\begin{figure}[t]
\includegraphics[width=7.5cm]{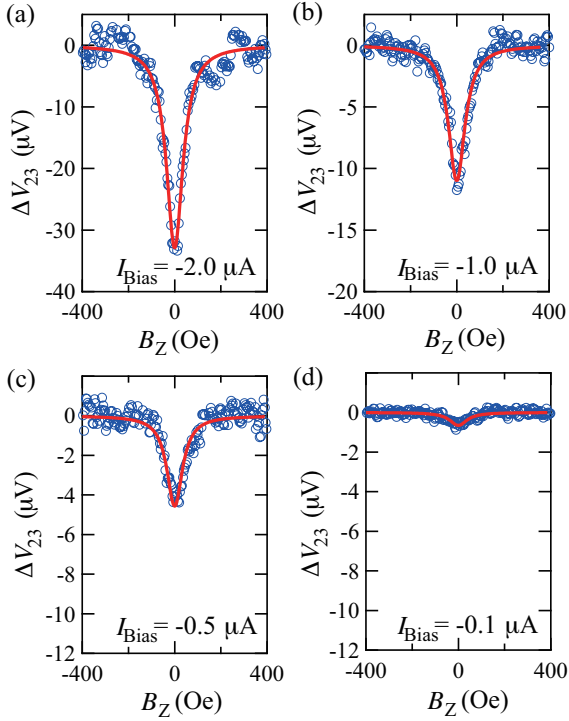}
\caption{(Color online) Room-temperature spin accumulation signals measured at $V_\text{G} =$ 8.0 V for various $I$$_\text{Bias} =$ (a) -2.0, (b) -1.0, (c) -0.5, and (d) -0.1 $\mu$A. The red curves are fitting results by the Lorentzian function.}
\end{figure} 

The three-terminal Hanle measurements were performed by a dc method with the current-voltage configuration shown in Fig. 2(a) at room temperature, where a small magnetic field perpendicular to the plane, $B_\text{Z}$, was applied after the magnetic moment of the contact 2 aligned parallel to the plane along the long axis of the contact. 
Figures 3(a)-(d) show $\Delta$$V_\text{23}-$$B_\text{Z}$ curves for $V_\text{G} =$ 8.0 V at various $I$$_\text{Bias} =$ -2.0, -1.0, -0.5, and -0.1 $\mu$A, respectively, at room temperature, where a quadratic background voltage depending on $B_\text{Z}$ is subtracted from the raw data. Here in this condition ($I$$_\text{Bias} <$ 0) the electrons are injected from the spin-polarized states of CoFe into the conduction band of Si, as shown in Fig. 2(a). When $B_\text{Z}$ increases from zero to $\pm$200 Oe, clear $\Delta$$V_\text{23}$ changes are observed for all $I$$_\text{Bias}$ even at room temperature. The presence of the changes in $\Delta$$V_\text{23}$ is caused by the depolarization of the accumulated spins.\cite{Jansen1} It should be noted that the magnitude of $\Delta$$V_\text{23}$ reaches $\sim$ 33 $\mu$V at $I$$_\text{Bias} =$ -2.0 $\mu$A, showing the resistance change in more than 15 $\Omega$. With decreasing $I$$_\text{Bias}$, the magnitude of $\Delta$$V_\text{23}$ is markedly reduced. This feature is consistent with the decrease in the injection of spin-polarized electrons.

We also examine $\Delta$$V_\text{23}-$$B_\text{Z}$ curves for various $V_\text{G}$ at a constant $I$$_\text{Bias}$ of -2.0 $\mu$A. Because of the large channel resistance and the presence of the large electrical noise, we could not obtain Hanle-like shapes in $V_\text{G} < 8.0$ V. At $V_\text{G} =$ 30 V, a relatively large change in $\Delta$$V_\text{23}$ of $\sim$ 29 $\mu$V is obtained in Fig. 4(a), but we find that the magnitude of $\Delta$$V_\text{23}$, $|\Delta$$V_\text{23}|$, is slightly reduced from $\sim$ 33 to $\sim$ 29 $\mu$V by applying $V_\text{G}$ from 8.0 to 30 V. 
Surprisingly, the magnitude of $\Delta$$V_\text{23}$ is further decreased to $\sim$12 $\mu$V when the gate voltage is further applied up to $V_\text{G} =$ 70 V in Fig. 4(b). Figure 4(c) displays $|\Delta$$V_\text{23}|$ vs $V_\text{G}$ for various $I$$_\text{Bias}$ at room temperature. With increasing $V_\text{G}$, $|\Delta$$V_\text{23}|$ is clearly decreased for each $I$$_\text{Bias}$. When we note again the $V_\text{G}$ dependence of $R_\text{Int}$ in the inset of Fig. 2(b), we find that the change of $R_\text{Int}$ with $V_\text{G}$ is not related to that of $|\Delta$$V_\text{23}|$. Thus, we focus on large changes in the characteristics of the Si channel, as shown in Fig. 2(d). We will discuss the detailed mechanism of the decrease in $|\Delta$$V_\text{23}|$ with increasing $V_\text{G}$ later. 
\begin{figure}
\includegraphics[width=8cm]{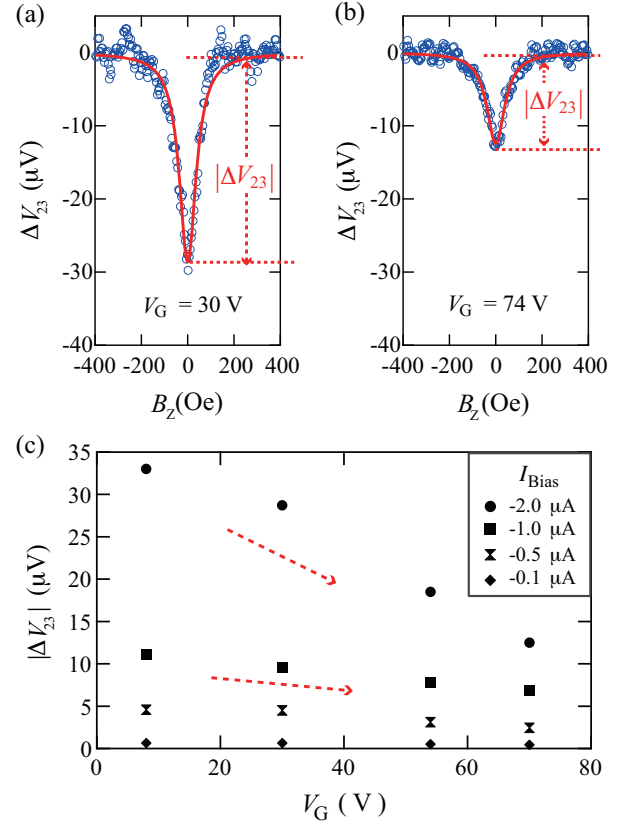}
\caption{(Color online) Room-temperature spin accumulation signals measured at (a) $V_\text{G} =$ 30 V and (b) = 74 V at $I$$_\text{Bias} =$ -2.0 $\mu$A. The red curves are fitting results by the Lorentzian function. (c) $|\Delta$$V_\text{23}|$ vs $V_\text{G}$ at various $I$$_\text{Bias}$ at room temperature.} 
\end{figure}

\section{Spin lifetime vs gate voltage}

The lower limit of spin lifetime ($\tau_\text{S}$) can be extracted from the obtained Hanle-effect curves.\cite{Jansen1} By using the Lorentzian function, $\Delta$$V_\text{23}$($B_\text{Z}$) $=$ $\Delta$$V_\text{23}(0)$/[1+($\omega_\text{L}$$\tau_\text{S}$)$^{2}$],\cite{Jansen2} the width of the Hanle-effect curve can be regarded as the value of $\tau_\text{S}$, where $\omega_\text{L} =$ $g\mu_\text{B}$$B_\text{Z}$/$\hbar$ is the Lamor frequency, $g$ is the electron $g$-factor ($g =$ 2), $\mu_\text{B}$ is the Bohr magneton. The fitting results were denoted by the red solid curves in Figs. 3(a)-(d) and Figs. 4(a) and 4(b). Figure 5 shows $\tau_\text{S}$ versus $V_\text{G}$ at room temperature, together with the change in $\rho_\text{Si}$ with $V_\text{G}$. The estimated $\tau_\text{S}$ values are ranging from 1.0 to 2.0 nsec. In general, the estimated $\tau_\text{S}$ values are affected by these doping elements and its doping density.\cite{Jansen1} For our devices, the spin accumulation is created in the P-doped channel beneath the contact with Sb $\delta$-doped interface. It should be noted that $\tau_\text{S}$ is almost constant whereas $\rho_\text{Si}$ largely changes more than one ordered of magnitude. 
This means that there is almost no correlation between $\tau_\text{S}$ and the change of the carrier density induced by $V_\text{G}$ in the Si channel, largely different from the dependence of the impurity doping density on the $\tau_\text{S}$ value presented in previous works.\cite{Jansen1,Jansen2,Jonker2} Therefore, at room temperature, we can regard the spin diffusion length ($\lambda_{\rm Si}$) as almost constant value for our devices. Assuming $D \sim$ 40 cm$^{2}$s$^{-1}$ ($n \sim$ 10$^{15}$cm$^{-3}$),\cite{Sze} we can also obtain $\lambda_{\rm Si} \sim$ 2.3 $\mu$m at room temperature by using the relationship of $\lambda_{\rm N} =$ $\sqrt{D\tau_\text{S}}$ ($\tau_\text{S} \sim$ 1.3 nsec). 
\begin{figure}
\includegraphics[width=8cm]{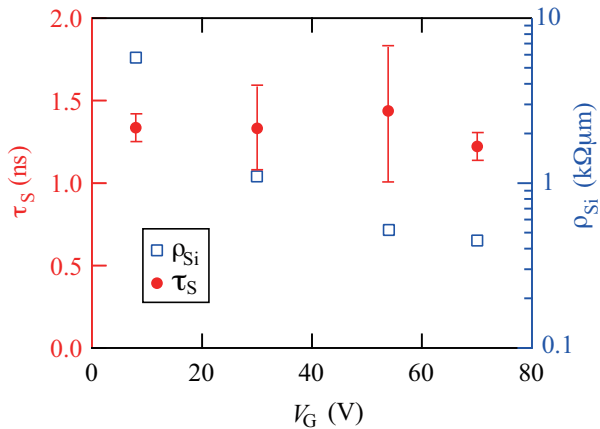}
\caption{(Color online) Spin lifetime ($\tau_\text{S}$) and channel resistivity ($\rho_\text{Si}$) as a function of $V_\text{G}$ at room temperature. }
\end{figure} 

\section{Magnitude of spin-accumulation signal}

When the current density ($J$) is assumed to be $|I_\text{Bias}|$/$A$, where $A$ is contact area (100 $\mu$m$^{2}$) of this device, the experimental spin resistance area-product (spin-$RA$) can roughly be estimated to be $\frac{|\Delta V_\text{23}|}{J}$ ($\Omega\mu m^{2}$). For example, the spin-$RA$ at $V_\text{G} =$ 8.0 V and $I_\text{Bias} =$ - 2.0 $\mu$A is roughly $\sim$ 1.65 k$\Omega$$\mu$m$^{2}$. On the basis of the simple spin diffusion model,\cite{Fert,Takahashi} the theoretical spin-$RA$ can be expressed as follows.\cite{Jansen1}

\begin{equation}
\text{Spin-}RA  = P^{2}\times \lambda_\text{Si} \times \rho_\text{Si} \times \frac{\lambda_\text{Si}}{d},
\end{equation}
where $P$ is the tunnel spin polarization, $\lambda_{\rm Si}$ and $\rho_\text{Si}$ are the spin diffusion length and resistivity of the Si channel used, respectively. $\frac{\lambda_\text{Si}}{d}$ is the geometrical factor of our devices,\cite{Jansen1} where $d$ is the channel thickness ($=$ 0.075 $\mu$m). By comparing Eq.(1) with experimental spin-$RA$ values, we can roughly estimate experimental $P$ values. Here $\lambda_{\rm Si} =$ 2.3 $\mu$m is assumed and $\rho_\text{Si}$ values at various $V_\text{G}$ are experimentally obtained in Fig. 5. In Fig. 6 we summarize the estimated $P$ as a function of $V_\text{G}$, together with the experimental spin-$RA$ at $I_\text{Bias} =$ - 2.0 $\mu$A. The estimated $P$ values are less than 0.2 for all $V_\text{G}$. Recently, the room-temperature spin polarization of epitaxial CoFe films on Si(111) was estimated to be less than 0.25 by nonlocal spin-signal measurements using metallic lateral spin valves.\cite{Oki2} The experimental tunnel spin polarization extracted here are smaller than $P =$ 0.25. From these considerations, at least, our experimental results can be understood within the framework of the theoretical spin diffusion model.\cite{Fert,Takahashi} We note that $P \sim$ 0.065 at $V_\text{G} =$ 8.0 V is relatively small compared with $P \sim$ 0.15 at other $V_\text{G}$. 

As shown in Fig. 5, $\rho_\text{Si}$ at $V_\text{G} =$ 8.0 V is one order of magnitude larger than those at other $V_\text{G}$. Since the resistance area product for our device is almost constant ($\sim$ 10$^{7}$ $\Omega$$\mu$m$^{2}$) despite changing $V_\text{G}$, we can understand that the spin injection efficiency for $V_\text{G} =$ 8.0 V is relatively low compared with that for other $V_\text{G}$ due to influence of the impedance mismatch problem.\cite{Jansen1,Rashba,Fert,Takahashi} We note that $|\Delta$$V_\text{23}|$ is linearly reduced in Fig. 4(c) despite the logarithmic decrease in $\rho_\text{Si}$ in Fig. 5.  Although we can tentatively speculate that the change in $|\Delta$$V_\text{23}|$ with $V_\text{G}$ originates from the change in $\rho_\text{Si}$ from Eq.(1), the correlation between $P$ and $\rho_\text{Si}$ with changing $V_\text{G}$ should be considered. From $V_\text{G} =$ 8 to 30 V, $P$ is enhanced from 0.065 to 0.14, leading to the enhancement in $|\Delta$$V_\text{23}|$, whereas $\rho_\text{Si}$ is logarithmically reduced, giving rise to the decrease in $|\Delta$$V_\text{23}|$. In this case, since the change in $\rho_\text{Si}$ is much larger that in $P$ from $V_\text{G} =$ 8 to 74 V, the change in $|\Delta$$V_\text{23}|$ is dominantly affected by the decrease in $\rho_\text{Si}$. Phenomenologically, even if the same $I$$_\text{Bias}$ is used for spin injection into the Si channel, the density of state in Si at the Fermi level can be varied by the application of $V_\text{G}$. As a result, the magnitude of spin accumulation ($\Delta \mu$) can be reduced.\cite{Ando4} 
\begin{figure}
\includegraphics[width=8cm]{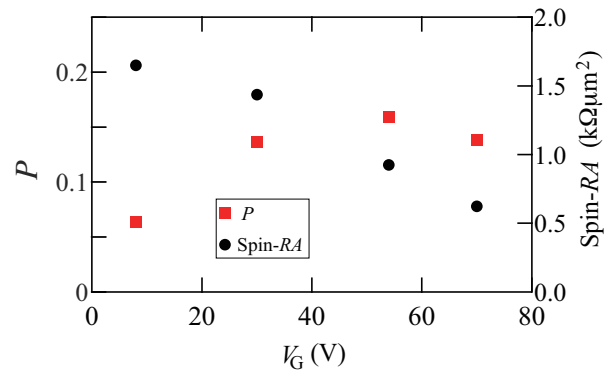}
\caption{(Color online) The estimated tunnel spin polarization ($P$) and the spin $RA$ as a function of $V_\text{G}$ at room temperature. }
\end{figure} 

We finally comment on a possible method to enhance the magnitude of spin signals at room temperature. Although relatively small $P$ of less than 0.2 has been obtained in this study (Fig. 6), an intrinsic spin polarization of our epitaxial CoFe electrodes is less than 0.25 at room temperature.\cite{Oki2} Thus, we can regard the spin injection efficiency as relatively high. Considering Eq.(1), we need to enhance $P$ value for the low-power consumption devices. Since we have already demonstrated highly ordered Heusler alloys, which have relatively high spin polarization intrinsically,\cite{Oki2,Hamaya3,KimuraHamaya}, grown on Si with keeping high-quality heterointerfaces, we will use these spin injectors and detectors to enhance the room-temperature spin signals for device applications. If we achieve nearly half-metallic spin injectors on Si, we would not have to consider the insertion of the relatively high interface resistance between Si and spin injector.\cite{Fert,Takahashi,KimuraHamaya,Ishikawa} As a result, the low parasitic resistance S/D contacts may simultaneously be realized.

\section{Conclusions}
We have studied room-temperature detection of spin accumulation in a nondegenerated Si channel with a doping density of $\sim$ 10$^{15}$cm$^{-3}$ using a MOSFET structure with a CoFe/$n^{+}$-Si contact. The observed spin accumulation signals can be modulated by the application of the gate voltage. We can interpret that the change in the spin-accumulation signals is attributed dominantly to the change in the resistivity of the Si channel, indicating reliable evidence for the spin injection into the nondegenerated Si channel at room temperature. We established a room-temperature spin injection technique for the nondegenerated Si channel without insulating tunnel barriers, indicating a technological progress for Si-based spintronic applications with gate electrodes.

\section{Acknowledgment}
KH would like to thank Dr. R. Jansen (AIST) for useful discussion. The authors thank Prof. A. Sakai (Osaka Univ.) and Dr. K. Izunome (Covalent Silicon Corporation) for their experimental supports. This work was partly supported by PRESTO from JST, STARC, and SCOPE from MIC. 


\end{document}